# A holographic mobile-based application for practicing pronunciation of basic English vocabulary for Spanish speaking children

Rebeca Cerezo[1], Vicente Calderón[2], Cristóbal Romero*[3]

[1]*Department of Psychology, University of Oviedo, 33009 Oviedo, Spain*

[2]*Department of English, Santiago Apóstol Infant and Primary School, 13760 Almuradiel, Spain*

[3]*Department of Computer Sciences and Numerical Analysis, University of Córdoba, 14071 Córdoba, Spain*

**Abstract**

This paper describes a holographic mobile-based application designed to help Spanish-speaking children to practice the pronunciation of basic English vocabulary words. The mastery of vocabulary is a fundamental step when learning a language but is often perceived as boring. Producing the correct pronunciation is frequently regarded as the most difficult and complex skill for new learners of English. In order to address this problem this research takes advantage of the power of multi-channel stimuli (sound, image and interaction) in a mobile-based hologram application in order to motivate students and improve their experience of practicing. We adapted the prize-winning HolograFX game and developed a new mobile application to help practice English pronunciation. A 3D holographic robot that acts as a virtual teacher interacts via voice with the children. To test the tool we carried out an experiment with 70 Spanish pre-school children divided into three classes, the control group using traditional methods such as images in books and on the blackboard, and two experimental groups using our drills and practice software. One experimental group used the mobile application without the holographic game and the other experimental group used the application with the holographic game. We performed pre-test and post-test performance assessments, a satisfaction survey and emotion analysis. The results are very promising. They show that the use of the holographic mobile-based application had a significant impact on the children's motivation. It also improved their performance compared to traditional methods used in the classroom.

**Keywords:** Holograms, Mobile Applications, Interactive Learning Environments, English Foreign Language, Learning at Early Age.

## 1.Introduction

Learning a foreign language such as English is centered on four basic skills: listening, reading, speaking and basic writing skills (Peregoy and Boilly, 2001). Normally, all of these skills are taught as a whole, but vocabulary learning is very important and is a prerequisite for language learning (Agca & Ozdemir, 2013). In fact, one of the components of mastering English as a Foreign Language (EFL) is vocabulary mastery (Nation, 2013), which means students being capable of understanding and using appropriate words (Zahedi & Abdi, 2012). Vocabulary, as Brown (2001) states, forms the building blocks of any language and because of its importance, vocabulary acquisition is receiving much more attention in Second Language (L2) pedagogy and research. Vocabulary can be classified as oral or written. Teaching and learning new vocabulary has traditionally focused primarily on definitions, but pronunciation is clearly an important factor in learning new words (Sweeting, 2016). Therefore teachers should facilitate this learning by not only explaining meanings but also modeling the pronunciation of these words. More specifically, the importance of children beginning to practice their pronunciation skills in an L2 at an early age has long been known to researchers and educators (Neri et al. 2008). It is much easier to introduce sound systems to children when they are usually simply copying what they hear, and building up mental generalizations based on their experiences (McMahon, 2002). Therefore, researchers and educators must devise effective ways of providing pronunciation training for young learners. A good example is the new concept of using of speech recognition technology in teaching language pronunciation to new learners (Liaw, 2014); a brief examination of the current literature demonstrates that this is a burgeoning field of research (Cavus, 2016).

Vocabulary acquisition is hard and requires more effort and time in L2 than L1 (Suwantarathip and Orawiwatnakul, 2015). In addition, many EFL learners find vocabulary learning boring, as they have to repeat and memorize unfamiliar words and spelling (Nguyen and Khuat, 2003). Teachers must find ways to help students learn English vocabulary in more motivating and interactive ways (Liu, 2014). Adult students are always interested in learning EFL at first. However, they gradually lose their interest in learning due to the difficulties they face when reading, communicating, and listening (Alkhalifah et al., 2012). Obviously, there are differences in teaching EFL to children compared to teaching adults or adolescents. Children are often more enthusiastic and lively as learners, but they also lose interest more quickly, they are less able to keep themselves motivated in tasks that they find difficult, and they do not have the same access as older learners to the metalanguage that teachers use to explain concepts (Cameron, 2001). Teaching a foreign language in infant education is a challenge (Flores and Corcoll, 2015) and English teachers who teach young children need to understand what it is like to be a child in order to be aware of the obstacles and difficulties that may be encountered in the process, and be ready for the challenge. Children have a natural instinct for play and fun, they demand a great deal of creativity and energy during interactions; if they are bored they will not pay attention, and they will not learn (Simona, 2015). Therefore, teachers of early-age learners have an important role to play in guiding children to English in a way which motivates them (Cameron, 2001). One such way is Computer-Assisted Language Learning (CALL) systems, which use games and multimedia in order to engage learners (Yip and Kwan, 2006). Previous research indicates that technology may increase students' motivation, especially in the case of children, and help them overcome previous difficulties (Alkhalifah et al., 2012; Finnsson, 2015; Wu, 2013). Teachers have the dual tasks of motivating and teaching, and one way of doing that may be by taking advantage of the huge potential of modern-day students' digital literacy and introducing technological innovations and computer games that enrich and encourage early-age learners to participate in educational activities in the classroom (Martins et al., 2015). Several studies suggest that computer applications and games seem to have a positive impact on EFL learning in school and pre-school (Finnsson, 2015) and are a key way to increase children's interest (Wu, 2013).

In this research we propose the use of a novel type of computer game developed with hologram technology, which is one of the most creative areas in the field of Information and Communications Technology (ICT) in learning environments (Ghuloum, 2010). The potential of audiovisual materials is already well known as they present a combination of sound, images and creative elements for learners to interact with (Squire, 2002). Holograms could enhance the educational process by bringing famous or fictional characters to life, so to speak, to talk about themselves or even add something as a classroom teacher, which could be interesting, motivating, and without doubt, an innovation in the teaching-learning process (Kalansooriya et al., 2015). With this same objective in mind, we adapted a holographic game to be used for educational purposes and developed a specific Android mobile application (app) to enable Spanish children to practice pronunciation of basic English vocabulary words. The app can also be used in a smartphone without the holographic game, in which case the image will be seen on screen rather than using holograms.

We chose practicing English pronunciation of basic vocabulary as the learning target because as well as being one of the first and most frequent skills taught to children learning English (Wu, 2013), it is also in the official Spanish academic curriculum for EFL at early-ages. We used a 3D holographic robot that acts as a virtual teacher. Children interact directly with the holographic robot through speech and see corresponding images as 3D holograms; aiming for them to make the association between the 3D image they see and the sounds they hear. A real teacher can use our holographic mobile-based application as an additional resource in the classroom to motivate children in a fun and entertaining way. Based on that, our research objective is to test the following hypotheses:

- **H1:** Children perform better when practicing pronunciation of basic English vocabulary words using a mobile-based application with a holographic game than using the traditional methods used in the classroom (images in books and on the board).

- **H2:** Children show a preference for using a mobile-based application with or without a holographic game for practicing the pronunciation of English vocabulary rather than the traditional methods used in the classroom.

The paper is arranged as follows: Firstly, we describe the related background. Then we present the adaptation of the holographic game together with the Android application, followed by the experimental results. We then discuss the results, including conclusions and lines for future research.

## 2.Background

A wide range of methods facilitate the acquisition of L2 vocabulary ranging from traditional instruction, dictionaries and pictures, to mobile applications and games (Chiu, 2013) (Nation, 2013). There are three categories of traditional techniques of teaching vocabulary (Gairns and Redman, 1986):

- Visual techniques. These techniques concern visual memory and consist of flashcards, photographs, pictures, blackboard drawings, wall charts, mimes and gestures. They are used in expressing words' meanings. These techniques are especially helpful in introducing certain parts of vocabulary such as real objects, places, professions, descriptions of people, action and activities.

- Verbal techniques. These include word lists, dictionary use, illustrative situations, synonyms and definitions, contrasts and opposites, scales and examples. These are very useful for illustrating abstract words.

- Translation. This is considered as an effective way to convey meaning. It helps to save time, especially in cases of teaching low frequency words. However, it is unproductive when teachers overuse translation.

Mobile learning (m-learning) refers to the use of mobile technologies for educational purposes. Such devices can offer multiple learning opportunities which may be spontaneous, informal, contextual, portable, ubiquitous, pervasive, and personal (Hwang et al., 2008). Numerous studies about the use of mobile technologies in education have been published in recent years (Hwang & Tsai, 2011). There are also hundreds of Mobile-Assisted Language Learning (MALL) publications in the last twenty years that use SMS/MMS, PDAs and smartphones (Burston, 2015). Nowadays there are an increasing number of mobile apps (software applications) for EFL available on the market (Pilar et al. 2013). These apps can be categorized into several groups according to their content: a) games, very often targeted towards children; b) app versions of dictionaries, handbooks and textbooks; c) apps providing vocabulary, grammar and/or pronunciation practice; d) the adaptation of online courses for mobile devices; and e) apps that provide the use of language in context, presented in a variety of ways such as podcasts, videos, films, and cartoons. In recent years, mobile learning has been increasingly used to facilitate ubiquitous language learning (Wang et al., 2017). However, vocabulary is the most often targeted skill making use of this type of technology (Suwantarathip and Orawiwatnakul, 2015). Mobile learning environments encourage students‘ curiosity and make learning activities more attractive and motivating (Agca and Ozdemir, 2013). There are a variety of specific mobile interactive vocabulary applications for improving EFL students‘ English pronunciation skills (Agusalim et al. 2014). Some of these apps use speech recognition engines (Cavus, 2016) which can recognize spoken words so that pronunciation errors can be easily identified and corrected. Speech recognition technology has also been used to support a group of elementary school children's EFL learning (Liaw, 2014).

Games for learning are an ideal mechanism for designing educational activities with children (Nacher et al. 2015). Different categories of games are used as educational tools (Freitas, 2006; StojkoviĤ and JerotijeviĤ, 2011) such as educational games, online games, serious games, simulations, structure games, vocabulary games and number games. Computer games have become a popular strategy for learning and there are an increasing number of examples of computer games developed specifically for young children and kindergarteners (Vangnes et al, 2012). Playing computer games has been found to be linked to a range of perceptual, cognitive, behavioral, affective, and motivational impacts and outcomes. Evidence of their effectiveness can be seen in existing results and data (Tobias, Fletcher, & Wind, 2014). Game playing is a good way to engage learners in language learning for a number of reasons (Prensky, 2001): games motivate players (to achieve goals), they gratify the ego (when winning), they are fun (through enjoyment and pleasure), and they spark the players' creativity (to solve the game). Games can focus on any of the various skills including grammar, listening, speaking, writing, reading, and pronunciation (Simona, 2015). The integration of learning content into a game context has the potential to produce better learning achievement and lower levels of anxiety in language learning (Hwang et al., 2017).

Magic games are a specific type of game that have been successfully used in education. Since children and adults are fascinated by magic tricks, this strategy for learning is effective and memorable (Carrasquillo, 2013). Magic is one of the oldest performing arts in the world in which audiences are entertained by staged tricks, effects or illusions of seemingly impossible or supernatural feats using natural means (Dunninger, 1987). Magic has been used to inspire children about science and technology for centuries (Curzon and McOwan, 2008). It is standard fare for Chemistry teachers to use 'magical potions' to engage children with chemistry. Similarly, Mathematics teachers use simple number based tricks to inspire wonder in mathematics. The use of magic tricks with children to assist in the development of cognitive, motor, speech, and psychosocial skills in therapeutic rehabilitation is well established (Spencer, 2012). Magic tricks have been used to enthuse children and to introduce formal methodological concepts in computer science (Curzon and McOwan, 2013).

In this research we adapted a Holographic magic show game to allow basic English vocabulary pronunciation to be practiced. A hologram is a 3D photographic image that appears to have depth. It can be defined as a three-dimensional record of the positive interference of laser light waves (Vincent, 2012). Holography began in the 1940s, when Dennis Gabor invented the hologram and won the Nobel Prize for the achievement (Gabor, 1948). Significant advances occurred when 3D Holographic Technology (3DHT) was created in 1962 by scientists in both the United States of America and the Soviet Union (Ghuloum, 2010). 3DHT operates by creating the illusion of three-dimensional imagery. Society has come to accept the idea of 3D images or holograms as being a part of our lives due to movies such as Star Wars and Star Trek (Sharton, 2010). Now 3D Holograms have broken out of the world of science fiction and fantasy and are also being used in advertising, entertainment, social networking, and more. However, as is the case with much technology, there are some barriers and disadvantages to 3DHT, for example the high cost of infrastructure, the need for a screening room with compatible lighting, and the lack of technical expertise. A way to avoid most of these problems is to use the illusion known as Pepper’s Ghost (Steinmeyer, 2013). This technique, which is used in theatre, haunted houses, and magic tricks, involves a large piece of glass or plastic film placed at an angle between the viewer and the scene. The glass or plastic reflects a room hidden from the viewer that is a mirror-image of the scene. Pepper's Ghost achieves the best effects by using a dark background with a single reflection. In this research we use the Pepper’s Ghost technique to create an affordable, accessible simulated holographic projection using a smartphone. Finally, to our knowledge, holograms have not yet been used in an educational environment with early age children and we have not found any magic show games that use holograms for teaching English to children as we propose here.

## 3. Adapting a Holographic Game and developing a mobile-based application

In order to create simulated holograms, we used a holographic game called HolograFX (http://holografxgame.com/) that allows users to perform magic shows using holograms on their smartphones. The game’s authors Mark Setteducati and Andre Armenante use the technology behind the classic Pepper's Ghost magic trick (Settembre, 2013). HolograFX was chosen as the 2013 Top Tech Toy at Toy Fair by the organizers of Gadget Show Live, the leading consumer tech event in the United Kingdom (http://www.toynews-online.biz/news/read/toy-fair-daily-gadget-show-names-holografx-as-best-tech-toy/031482). HolograFX (see Figure 1) uses a stage with a transparent reflective plastic screen in the middle and on one side a Smartphone holder for hiding the mobile. Users perform actions and tricks by using accessories on the opposite side of the smartphone which projects a hologram on the screen to support the show. HolograFX does not include a smartphone with the game components so one is required on which the specific HolograFX app can be installed. This app is not aimed at education and contains prerecorded videos, which combine holograms with tricks such as teleportation, and objects appearing and disappearing.

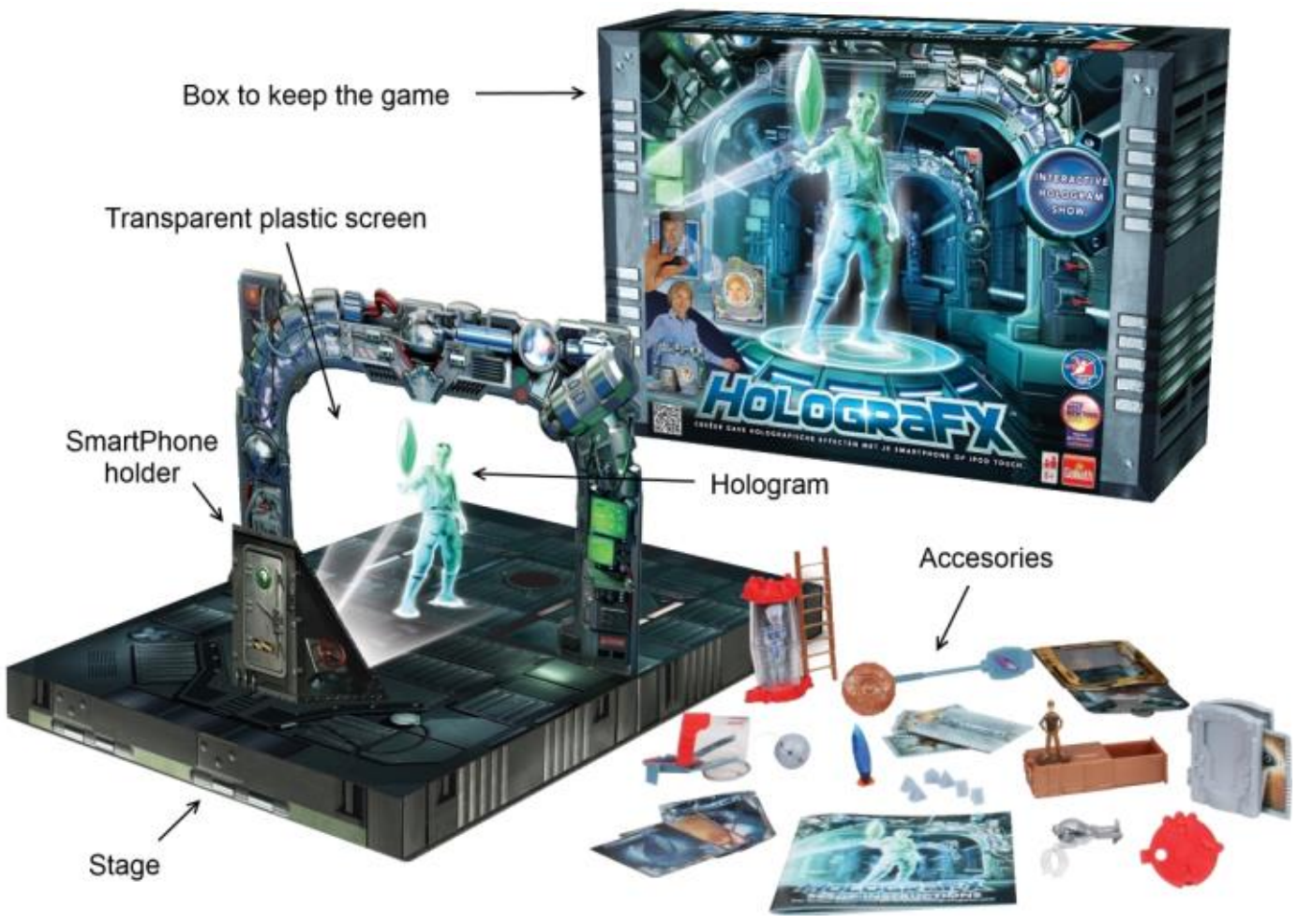

Figure 1: HolograFX game.

In order to adapt the game to allow children to practice and improve their pronunciation of basic English vocabulary through 3DHT we developed a completely new Android app (a demo is free to download from Google Play for Education https://play.google.com/store/apps/details?id=com.cromero.hi2). The app is aimed at practicing the pronunciation of basic English vocabulary such as animals, plants, vehicles, clothes, etc. Although our app uses a similar interface to the original HolograFX app, it is a new application with completely different functionality (see Figure 2). The first image in Figure 2 shows the main menu, with the three options: learn/revise, practice, and evaluate English vocabulary. We used a 3D robot, called *Arturito*, as the virtual teacher (see the 2nd image in Figure 2), and animated 3D images that represent the corresponding words (see the 3rd image in Figure 2) as holograms. For each vocabulary word our app has a black background video file with an animation of the corresponding 3D object and a sound file with its correct English pronunciation. We also included a scoreboard to give the teacher a summary of the children's answers (see the 4th image in Figure 2).

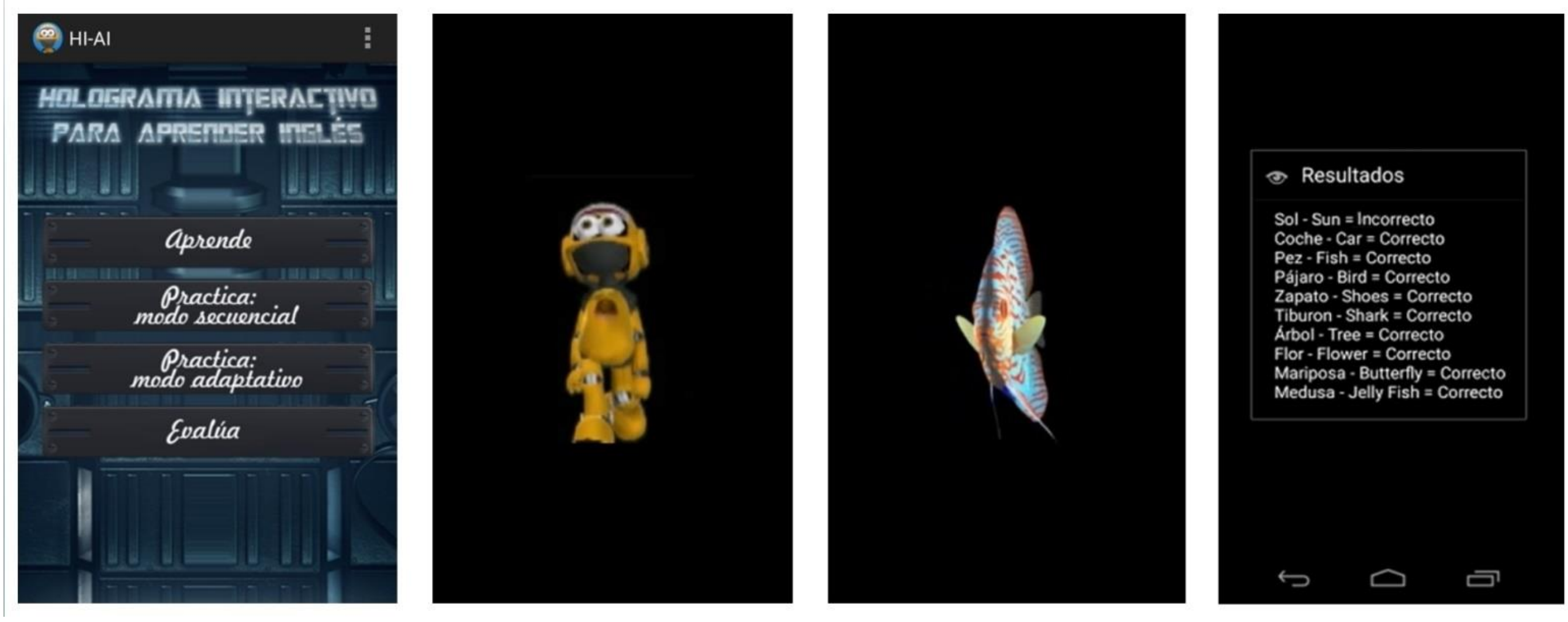

Figure 2: Screenshots of the Android application for practicing pronunciation of basic English vocabulary words.

The app has 4 main options or function modes (see the first image in Figure 2):

- **Learn/Revise (*Aprende*)** mode**.** *Arturito* introduces each term in Spanish (see image three in Figure 2), then repeats each term in English. In this study we used this mode for revising the vocabulary.
- **Practice: sequential mode (*Practica: modo secuencial*).** In Spanish, *Arturito* asks a student how to say each term in English in sequential order, and waits for a reply. For each attempt, he says whether the student's answer is correct or not.
- **Practice: adaptive/random mode (*Practica: modo adaptativo/aleatorio*).** In Spanish, *Arturito* asks a student how to say the term in English, in this mode the questions do not follow a sequential order, but are presented in a random or adaptive order. In this study we only used the random order. For each attempt, *Arturito* says whether the student's answer is correct or not.
- **Evaluate (*Evalua*) mode.** In Spanish, *Arturito* asks a student how to say a word in English. He does not tell each student whether their answer is correct or incorrect. Finally, a scoreboard (see the last image in Figure 2) is displayed with the children's results for each word: Correct (Correcto) or Incorrect (Incorrecto).

We added an animation of *Arturito* singing and dancing at the end of each mode to encourage the children to pay attention and listen to him model the pronunciation of the target words. Our goal was to keep the children engaged and keen to complete each mode and to complete the tasks that would allow them to see what *Arturito* would do at the end, presenting them with a fun, game-based learning environment, encouraging them to work towards a goal without worrying about making mistakes. The children are not being taught to speak like robots; the pronunciation of the words was recorded by a native English speaker and a subtle robot filter was applied to the audio to make it slightly robot-like.

We chose a robot-like figure to act as a teacher because robots are becoming an integral part of our society and have shown great potential when used as an educational technology (Mubin, 2013). There are several previous studies that show that interactive robots can be more effective than learning from classical on-screen media (Kose-Bagci, 2009; Leyzberg, 2012). Young children performed better in post-learning tests and had more interest when language learning took place with the help of a robot compared to audiotapes and books (Han, 2008). We believe that children would like to have a Holographic robot that acts as teacher to practice word pronunciation. We chose the specific shape and the name *Arturito* as an *homage* to the two famous Star Wars robots R2D2 and C3PO. *Arturito* is a similar shape to the humanoid robot C3PO and the name comes from the pronunciation of R2D2 (Ar-Two-Di-Two), which when said quickly sounds similar to *Arturito* in Spanish.

All interaction between *Arturito* (hologram) and the students is via sound and voice (see Figure 3). We used the Android Speech API to access the speech recognition service (https://developer.android.com/reference/android/speech/SpeechRecognizer.html). A native British English speaker had recorded the correct pronunciation of the vocabulary words, which were stored as voice files. In this case we used British standards for pronunciation. However, this API uses a speech recognition engine that can understand a wide variety of languages that can be easily changed. When a student speaks or replies to *Arturito*, their audio is automatically translated into text and compared to the corresponding vocabulary words in order to check if it is correct or incorrect through the Speech Recognition API. The app plays a short tone just after asking for a word to be pronounced to let the children to know when they have to speak. The captured audio from the smartphone microphone is sent to the Speech API service and a text transcription is received, in our case, a word. The app then compares the returned word (what the children said) against the correct word (the word corresponding to the Holographic image *Arturito* displayed) in order to determine if the pronunciation was correct or not.

During the learning experience, the students do not look at the screen of the smartphone; they see the hologram as though it is on the other side of the plastic element (see Figure 3). The children believe that they are talking directly to *Arturito* and they do not notice that they are really interacting via voice and sound with the hidden Smartphone on their side of the stage. The app can be also used without the HolograFX game. In that case, the teacher or student must hold the smartphone in their hand and look at the screen. This means direct interaction with a smartphone device via voice and sound as *Arturito* and the images appear directly on the screen of the smartphone instead of as a hologram.

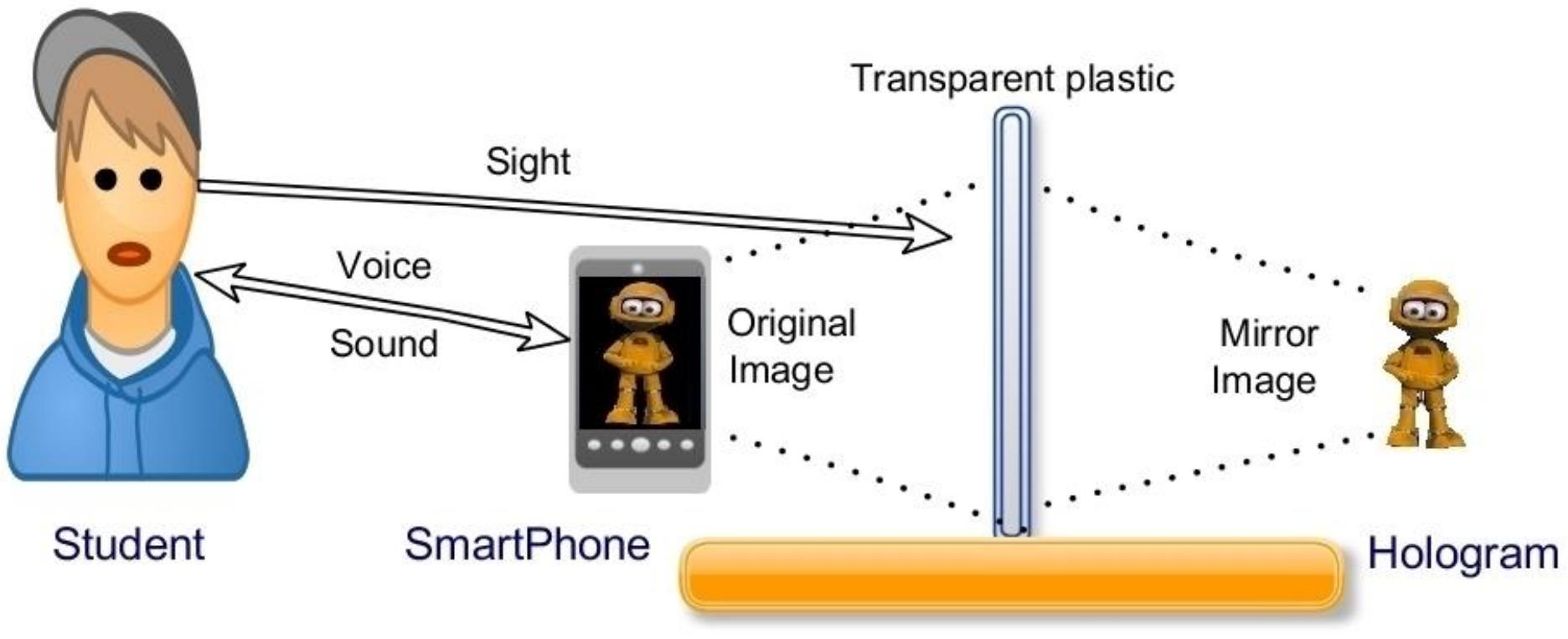

Figure 3: Student-Hologram interaction.

We also made some physical modifications to the structure of the original HolograFX game (see Figure 4) in order to improve it for use in a classroom setting with children. We added the following:

- A large black box to cover the hologram side of the stage to improve the visibility of the hologram in daylight.
- A picture of a traditional microphone at the back of the holder where the Smartphone is hidden so that children know that they have to speak towards that point.
- A miniature wooden desk/table as an accessory to simulate a traditional teacher's desk above which the holograms appear.

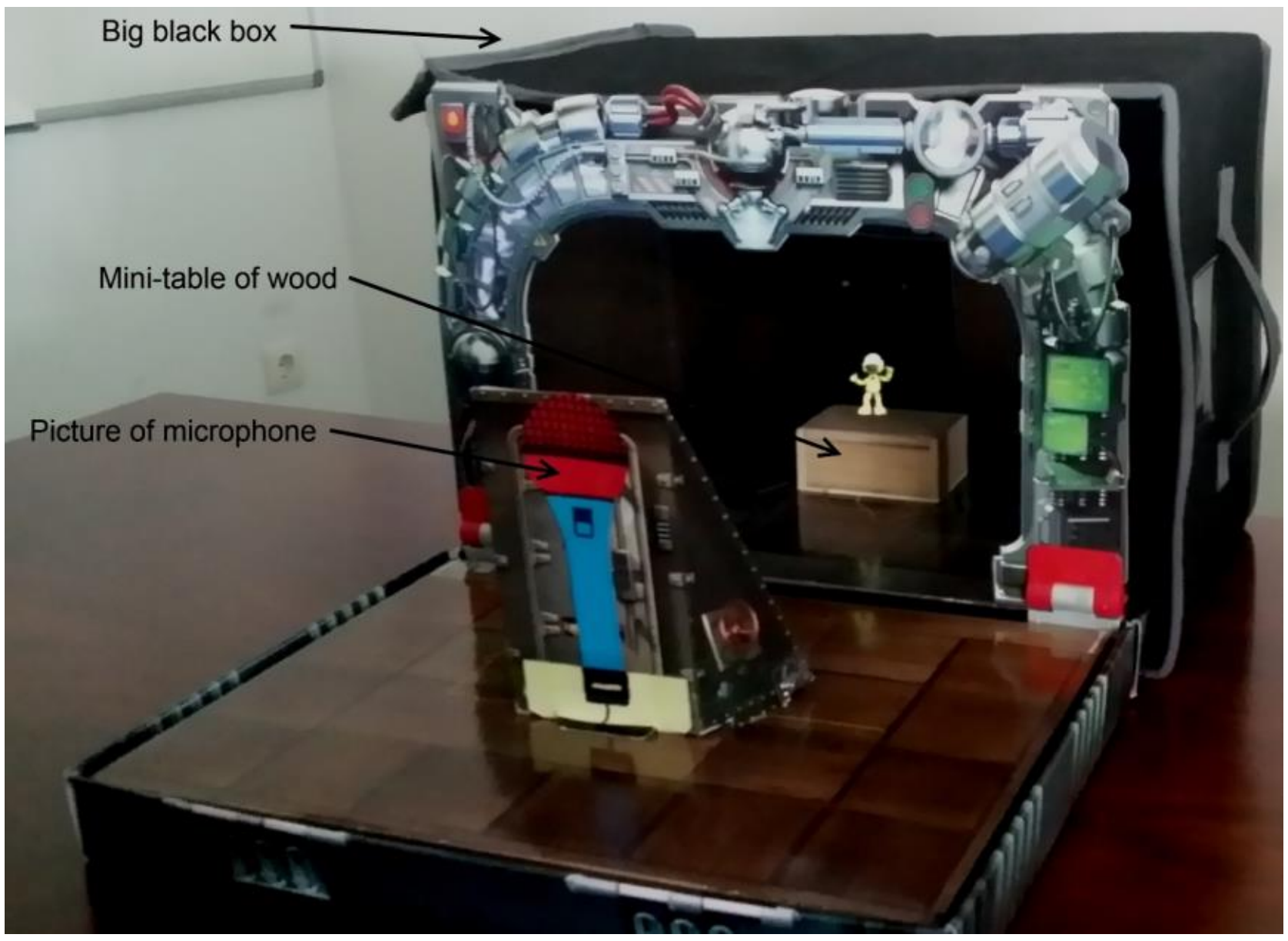

Figure 4: Modifications in the structure of the holographic game.

Finally, we also tried to maintain the original magical feeling that the interaction of the hologram and accessories produced in the original game. To do that, we used a miniature desk accessory, into which we could put small cards (see Figure 5) with pictures of both *Arturito* and the vocabulary words. The children can be shown the small cards being put into the mini-desk and then "magically" becoming real from inside it. The teacher can easily add these cards by capturing the images from the smartphone and using a computer to resize and print them.

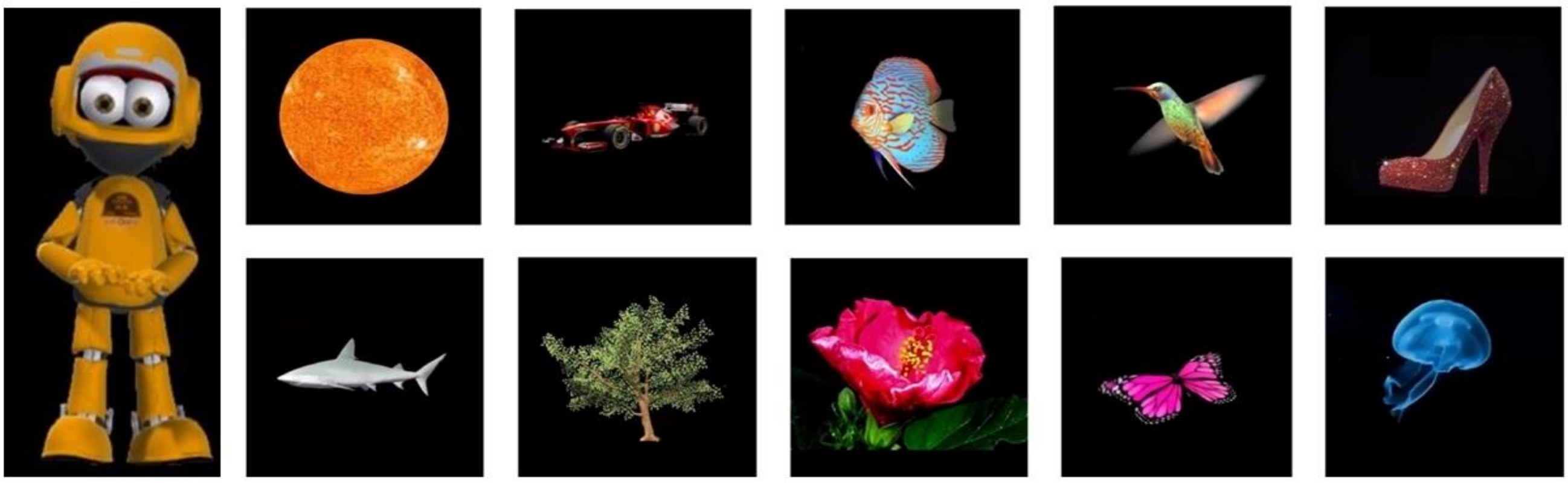

Figure 5: Examples of small cards used as accessories to put into the mini-desk.

## 4.Method

We performed an experiment with 70 pre-school children divided into three classes that were randomly assigned to one of the different groups. The first class was the control group using traditional methods such as images in books, on cards, and on the board, while the other two were experimental groups using the mobile application. One experimental group used the application without 3DHT and the other experimental group used the application with 3DHT. In both cases the application used the same types of interactions but either with or without the hologram element of the game. We conducted a pre-post-test design along with a final satisfaction survey and an image emotion analysis.

### *4.1. Participants*

Our participants were 70 pre-school Spanish children from a public school in Spain divided into three classes with a similar level of performance in English. The children were between 4 and 5 years old and in their second year of the pre-school period. The experiment was performed during the third term in the foreign language class in the academic year 2015-2016. The objective was to assess different teaching methodologies for practicing the pronunciation of basic English vocabulary words learned during the first and second terms.

The three classes had the same teacher who was also in charge of implementing the experiment. Although the study had not recruited students on a random basis but instead used existing classes, the teacher randomly divided each class into four subgroups (about 6 children each), and each class was randomly assigned to either control or experimental groups as follows:

- Control Group (CG): Class 1 with 24 children (13 boys and 11 girls) divided into four subgroups of 6 children each. They used traditional images in books, on cards and on the board.
- Experimental Group 1 (EG1): Class 2 with 23 children (10 boys and 13 girls) divided into four subgroups (three subgroups of 6 and one of 5). They used the app without the holographic game.
- Experimental Group 2 (EG2): Class 3 with 23 children (12 boys and 11 girls) divided into four subgroups (three subgroups of 6 and one of 5). They used the app with the holographic game.

### *4.2 Procedure*

The experiment was about practicing pronunciation of previously learned English words. In this study, the children practiced pronunciation of individual words although they had been taught these words in appropriate contexts during the first and second terms. In the third term, we wanted them to practice the pronunciation of these words in order to evaluate different teaching methods.
We selected 20 random words of differing lengths (one-, two- and three-syllable concrete nouns) from different topics covered in the classroom (animals, plants, vehicles, clothes and space) (see Table 1). The vocabulary was selected with the agreement of two infant education teachers, based on texts and reading books used at that educational level, trying to avoid altering the academic syllabus or course content. Two different versions of this vocabulary were used. Both contained pairs of words from the same topic which were closely related and had a similar level of difficulty so that they could be used in the pre- and post-testing. The full vocabulary list (version 1 plus version 2) was used for revising and practicing (see Vocabulary v1 and v2 in Table 1). Version 1 vocabulary was used for pre-testing and version 2 vocabulary was used for post-testing.

*Table 1.* Vocabulary used in the experiments.

| **Vocabulary v1** | **Vocabulary v2** |
|---|---|
| Sun | Moon |
| Car | Bus |
| Fish | Seal |
| Bird | Duck |
| Shoes | Boots |
| Shark | Crab |
| Tree | Grass |
| Flower | Cactus |
| Butterfly | Dragonfly |
| JellyFish | Starfish |

We used quantitative data to test our initial hypotheses. In particular, we used a performance pre-post-test for testing H1 and a satisfaction survey and picture emotion analysis for testing H2. Figure 6 gives an overview of the study’s five-week experimental procedure.

Figure 6: Diagram of experimental procedure.

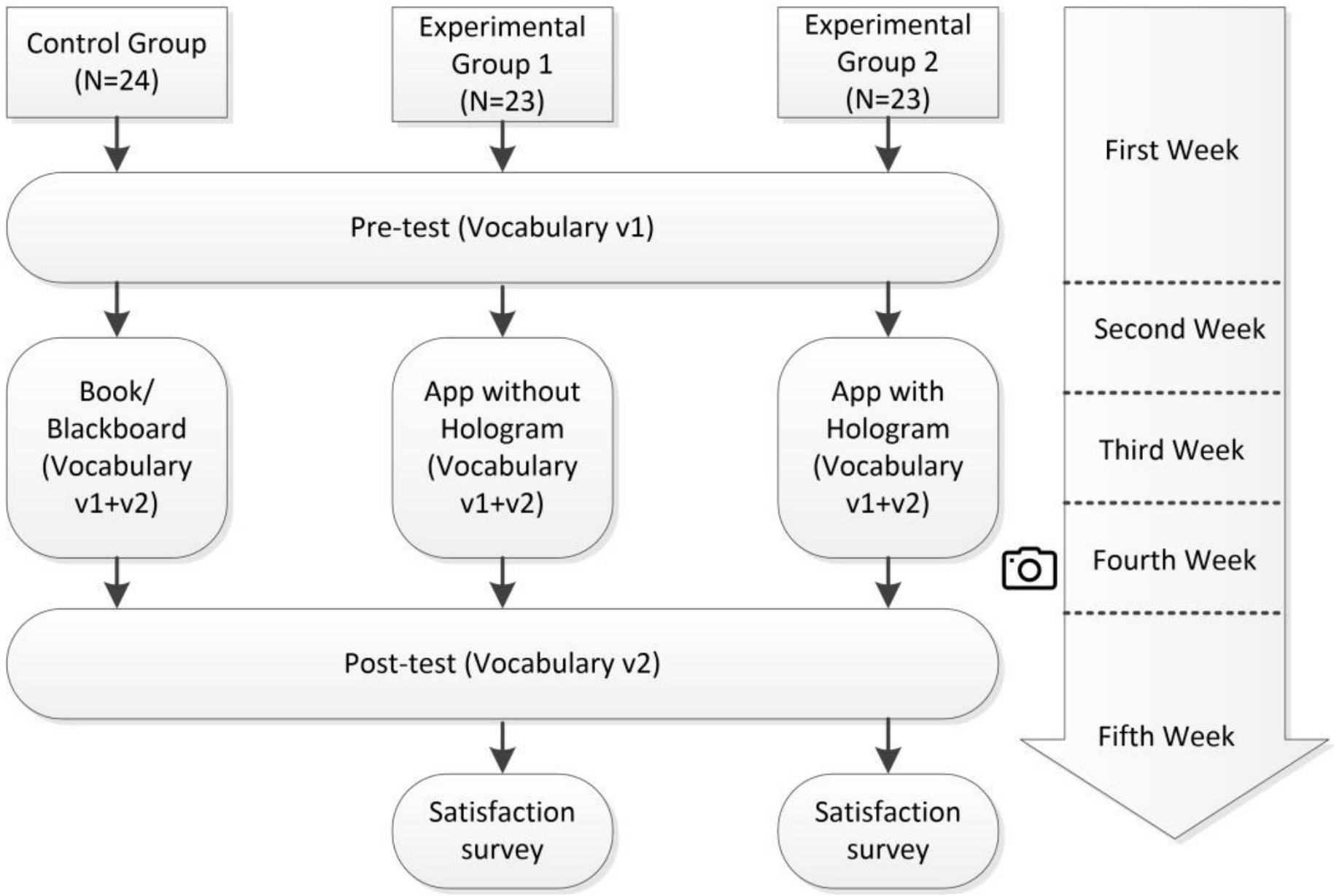

At stage 1, all students completed a pre-test to evaluate their prior knowledge about the pronunciation of the words by using vocabulary version 1. At stage 2, the children learned and practiced the pronunciation using the full vocabulary list (version 1 plus version 2) in 30-minute sessions once a week for 3 weeks, with each group using a different approach. All participants interacted orally (listening and speaking) during the experiment. During the revision/learn mode, the children looked at drawings, pictures or 3D image holograms that represented each word and listened to the correct English pronunciation of each word modeled by the instructor or the *Arturito* hologram. During the practice mode, they also tried to correctly pronounce each word in English that they saw in the images.

Experimental groups EG1 and EG2 used the app with or without the Hologram. In this research, we wanted to study two different experimental conditions, a drill and skill – mobile based app that closely mirrors the types of interaction available with the hologram but without the hologram element; and the hologram – holographic game (also mobile-based but with the addition of the holographic technology). We included the intermediate mobile based app condition to provide us with information about the role of the holographic technology.

The first week was spent introducing the mobile-based application by using the revise/learn mode in which *Arturito* introduced all the words from the full vocabulary list. The following two weeks were spent practicing the vocabulary by using the sequential and random practice modes. In the first week *Arturito* asked for words from the full vocabulary list in sequential order and the second week in random order. All of the 2D or 3D images representing the full vocabulary list appeared in sequential or random order. During each of these three sessions, each child interacted with the application in turn for about 5 minutes (about 15 minutes in total during the three weeks) while the others observed and remained quiet. The experiment was presented to the children as a simple game as follows: "You have to interact with a virtual teacher called *Arturito* to practice pronunciation of English vocabulary words. Try to pronounce all the words we put into the box correctly, *Arturito* will ask you by showing you a picture. If you do it right you will be able to see *Arturito* dancing afterwards."

During the same period, the CG used images of the same words from books and on the board. The sequence of revising and practicing was the same: the first week was used to revise the full vocabulary. The teacher used the board with pictures that represented each word from the full vocabulary list. The teacher pointed to each picture and pronounced it in English while students listened. The following two weeks were spent sequentially and randomly practicing the vocabulary by using a book with images of the vocabulary (one image per page). In the first week, the teacher asked for words from the full vocabulary list using images from the book in sequential order, starting at the first page and continuing to the last page of the book. The teacher pointed at each image in the book, asking the children for the corresponding word. The following week, the teacher asked for items from the full vocabulary list randomly. The teacher opened the book on a random page and pointed to the image each time, asking for the word. If a student had already seen that image, the teacher continued opening pages at random until a new word was shown.

The same teacher conducted the different sessions in the four subgroups from each class, while an additional teacher was with the rest of the children in another classroom. The experiment was carried out in a classroom next to the children's usual daily classroom, and in a similar environment in order to ensure that the children did not feel awkward being out of their usual place of learning. Another additional researcher (as a non-participant observer) took pictures of all of the children during the final practice session (fourth week) in order to observe students' emotions during classes with and without holograms. The aim of this was to have a quantitative measure of the emotions the children expressed, in order to test the second hypothesis, in addition to the satisfaction survey.

At stage 3, all students completed a post-test to evaluate their post knowledge using vocabulary version 2. Finally, EG1 and EG2 completed a satisfaction survey about their experience using the app with or without the hologram in contrast to the traditional methods. All the data about the experiment were directly managed and collected by the researchers.

### *4.3 Data Analysis and Results*

#### *4.3.1. Children's pronunciation performance*

In order to test the 1st research question *-do children perform better when practicing pronunciation of basic English vocabulary words using a mobile-based application with or without a holographic game than using traditional methods?-* We measured the performance of the students using a pre-post-test experimental design. The pre-test and post-test performance assessment of the pronunciation of words was done individually one week before the first session and one week after the last experimental session. The assessment consisted of a simple 10 playing-card (flashcards without text) question test, given non-sequentially where every correct pronunciation counted as 1 point. Each child's final score was an integrated value out of 10. From a teaching and educational psychology point of view, it seems to be an objective, operational way to measure the children's performance, due to their very young age. Both the pre-test and post-test were exactly the same for the three groups (CG, EG1 and EG2) and were carried out by the same teacher who carried out the experimental sessions.

The data produced by the three experiments were analyzed to verify that there were no values outside of the scale, missing values, or parameters that indicated a clear non-normal distribution; a Levene's test on the samples from the three groups was conducted, and the homogeneity test showed no significant difference ($F = 0.269$, $p > 0.05$). The descriptive statistics about learning performance are given in Table 2.

*Table 2*. Descriptive statistics for pre-test and post-test.

| Group | Variable | Mean | SD |
|---|---|---|---|
| CG (N=24) | Pre-test | 3.95 | 0.95 |
| | Post-test | 7.00 | 1.02 |
| EG1 (N=23) | Pre-test | 4.04 | 1.33 |
| | Post-test | 7.52 | 1.16 |
| EG2 (N=23) | Pre-test | 4.13 | 1.05 |
| | Post-test | 8.08 | 1.12 |

*p<0.05

To start with the inter-group comparison, an analysis of variances (ANOVA) on the pre-test was performed in order to test whether there were significant differences between the three groups. Results of the one-way ANOVA on pre-test scores are shown in Table 3. This analysis ($F=0.13$, $p \geq 0.05$) indicated that there was no significant difference in the three group's levels of prior knowledge before the experimental teaching activities began.

*Table 3*. ANOVA results of the pronunciation achievement for the pre-test.

| Variable | Source | Sum of Squares | Degrees of freedom | Mean Square | F statistic |
|---|---|---|---|---|---|
| Pre-test | Between groups | 0.34 | 2 | 0.17 | 0.13* |
| | Within groups | 84.52 | 67 | 1.26 | |
| | Total | 84.87 | 69 | | |

*p<0.05

We also calculated the Cohen's d about the strength of the relationship between the CG and the two experimental groups based on the standard difference between the post-test means. Cohen's d between CG and EG1 (0.47) indicated that the effect size was moderate and Cohen's d between CG and EG2 (1.00) indicated that the effect size was large.

Next, we conducted an analysis of covariance (ANCOVA), using the pre-test levels as covariates in order to statistically control the effect of the initial level of knowledge on the post-test level even though there were no statistically significant differences, this allowed us to gather more accurate information about the effect of the holographic technology. The results of ANCOVA are shown in Table 4

*Table 4.* ANCOVA results of the pronunciation achievement for the post-test

| **Source** | **Sum of Squares** | **Degrees of freedom** | **Mean Square** | **F statistic** | **Post Hoc** |
|---|---|---|---|---|---|
| Adjusted means | 10.56 | 2 | 5.28 | 12.397* | EG1 > CG<br>EG2 > CG<br>EG2 > EG1 |
| Adjusted error | 28.11 | 66 | 0.426 | | |
| Adjusted total | 38.67 | 68 | | | |

*p<0.05

The results of the ANCOVA (see Table 4) showed statistically significant differences between groups, ($F_{(2, 66)} = 12.397$, $p<.001$, $\eta_p^2 = .273$). These results also provide information about the autoregressive effect of the variable, the capacity of the pre-test levels to predict the post-test levels of pronunciation performance, $F_{(1, 66)} = 53.455$, $p< .001$, $\eta_p^2 = .655$. The results suggest that previous knowledge of pronunciation of English words is a variable that is not easy to change and very strongly influenced by its high pre-test level; in this sense, the level of previous knowledge could be obscuring the true potential of the hologram technique. Nevertheless, it is revealing to check the differences between the three groups. The comparison shows that the post-test scores of the CG and EG1, and CG and EG2 are significantly different $MD_{(CG-EG1)} = -.454$, $p< .05$; $MD_{(CG-EG2)} = -.950$, $p< .001$ , and that the post-test scores of EG1 and EG2 are also significantly different ($MD_{(EG1-EG2)} = -.496$, $p< .05$). In short, these results show that although all three groups improved their pronunciation of English vocabulary words, EG2 scored significantly higher than EG1 and the CG.

Additionally, the results from the intra-groups analysis better illustrate the inter-group differences shown. As seen in the descriptive statistics, the initial level in the pre-test is about 4 points in all groups. This shows that the children already had some knowledge of the correct pronunciation of the words in the pre-test, as they had seen them in the first and second terms, even so, the children’s pronunciation post-test was better in all groups. The scores in the CG ($t=23.87$, $p<0.05$), EG1 ($t=21.10$, $p<0.05$) and EG2 ($t=29.73$, $p<0.05$) were statistically significantly different before and after the experiment. However, this change was higher in the EG2 group (+3.95) than in the EG1 (+3.48) and CG group (+3.05) (See Table 5).

The results shown could be presumably be attributed to the holographic component and its capacity to increase student’s motivation and encourage learning in the classroom over the more common mobile-based applications or traditional methods using images in books and on the blackboard.

*Table 5.* Paired-sample t-test and increased performance for pre-test and post-test

| Group | t | Increased Performance |
|---|---|---|
| CG (N=24) | 23.87* | + 30.5% |
| EG1 (N=23) | 21.10* | + 34.8% |
| EG2 (N=23) | 29.73* | + 39.5% |

*4.3.2 Children 's satisfaction and emotions*

In order to test the 2nd research question *-children show that they prefer using a mobile-based application with or without a holographic game for practicing oral English vocabulary rather than the traditional methods used in the classroom-* we used two methodologies, a satisfaction survey and an emotion analysis.

We developed a specific satisfaction survey to ask children from both EG groups about their experience with the app. First, we asked them about their satisfaction compared to regular teaching methods (See Dimension A) because of the well-known importance of motivation in every kind of learning (Schunk & Zimmerman, 2012; Schunk, Meece, & Pintrich, 2012). We only asked children from the experimental groups because they experience books and blackboard learning every day and they also used the app whereas the control group only used the traditional methods with images in books and on the board. Following that we wanted to focus on the hypothetical differences in satisfaction with holographic technology between the two EGs. We asked the children in the EGs about the learning experience specifically to check if there were any differences in terms of satisfaction between using the mobile app with and without holographic technology (See Dimension B). To do this we carried out a very specific, simple ad hoc satisfaction survey due to the participants' young age and self-reporting skills; it had a Cronbach's alpha of 0.88. The survey consisted of six questions each with a 3-point Likert Scale (0=A little, 1=Some, 2=A lot) about two dimensions: learning experience satisfaction in comparison with traditional methodology and general experience satisfaction. In this satisfaction survey, we used only positive statements and as simple a scale as possible in order to make the questions easier to understand. It is well known from Piaget's theory of cognitive development that negatively formulated questions can pose serious problems in survey research with early age children (Wadsworth, 1996) and even adults (Suárez-Alvarez, et al., 2018). In fact, there are several studies that confirm a negative effect on response reliability when children answer negatively formulated questions (Borgers et al., 2000) because they have not yet developed the necessary formal thinking to understand logical negations (Borgers et al., 2004).

Dimension A. *How much they like Arturito compared to traditional methods*

A1. English classes are better because they are with *Arturito* instead of books and the board.

A2. *Arturito* is cooler than the books and the board.

A3. I want *Arturito* to teach me more stuff instead of using books and the board.

Dimension B. *How much they like Arturito*

B1. I like *Arturito*

B2. *Arturito* is funny

B3. I would like *Arturito* to teach me for a longer time.

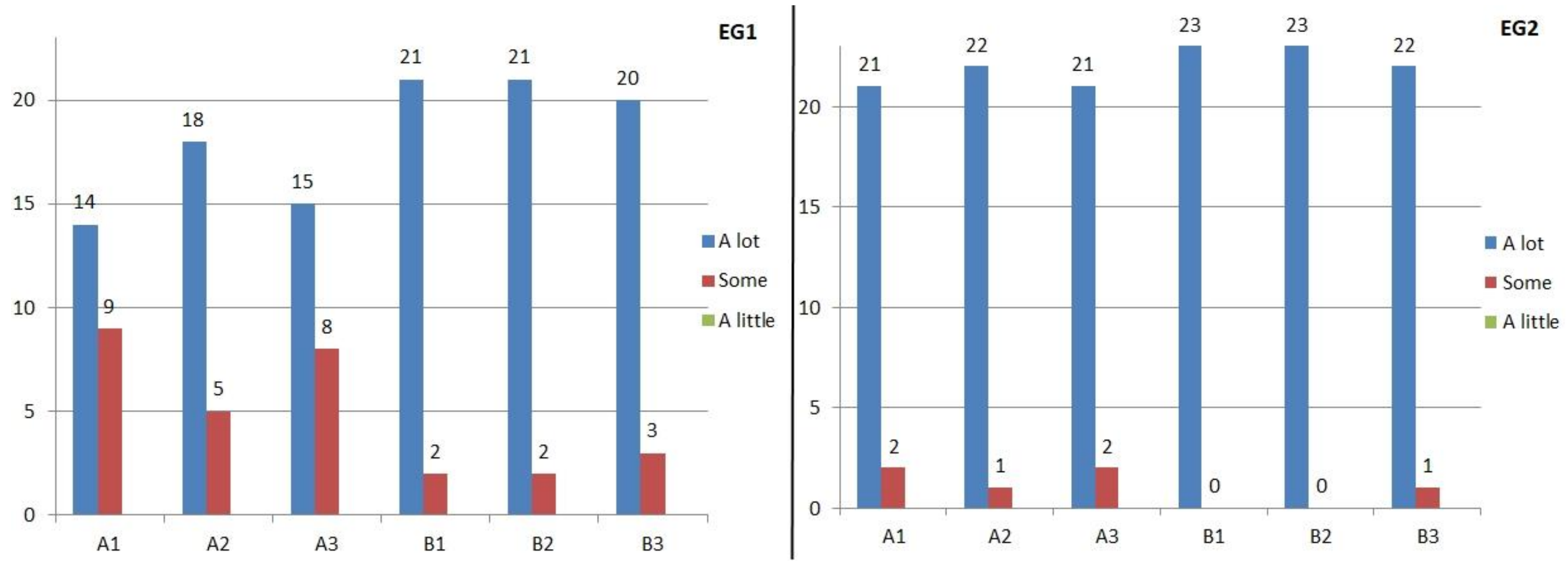

Figure 7: Frequency analysis of EG1 and EG2 answers´ in satisfaction survey.

The teacher asked each student the six questions, giving them the 3 answer options and recording their answers. Figure 7 shows that EG2 gave higher values than the EG1 group in every item, especially in dimension B questions, testing the differences in terms of satisfaction between mobile and holographic technology. In EG2, almost all the children (between 21 and 23) gave the highest value on the scale for all questions and only 1 or 2 children answered “Some” in some items (A1, A2, A3 and B3). In EG1, between 14 and 21 children responded with the highest value “A lot” to all the questions and between 2 and 9 children answered “Some” to the questions. Therefore it seems that satisfaction was very high in general, but slightly higher in EG2 than in EG1.

Table 5 shows the descriptive statistics and t-test from the satisfaction survey in EG1 and EG2. The results about preference for *Arturito* over traditional methodology are higher in EG2 (Dimension A: M=1.92, SD=0.25) than EG1 (Dimension A: M=1.69, SD=0.46) (see Table 6). In this dimension, question A2 gave the highest mean score in both groups, providing evidence that children think that *Arturito* is cooler than traditional methods. The average sense of satisfaction with *Arturito* was very high in both groups although it was higher in EG2 (Dimension B: M=1.98, SD=0.06) than EG1 (Dimension B: M=1.89, SD=0.29). In this dimension, questions B1 and B2 gave the highest mean scores in both groups, providing evidence that children liked *Arturito*. In fact, all of the children in EG2 answered these two questions with the highest value on the scale.

We also performed a t-test (see Table 6) in order to examine whether there were differences between the mean scores in each question. We found that there were no significant differences between the EG1 and EG2 scores in questions B1 (t=-1.41, p<0.05), B2 (t=-1.41, p<0.05), B3 (t=-0.99, p<0.05) and A2 (t=-1.70, p<0.05), but there were significant differences in A1 (t=-2.42, p<0.05) and A3 (t=-2.11, p<0.05). So in general, both groups expressed the idea that they liked *Arturito* and that he is cooler than traditional methods without significant differences. But children in EG2 were statistically significantly more satisfied than children in EG1 with the holographic *Arturito* compared to just using images in books and on the board.

Table 6. Descriptive statistics and t-test for satisfaction survey in EG1 and EG2.

| **Dimension** | **Question** | **Group** | **Mean** | **SD** | **t** |
|---|---|---|---|---|---|
| **A** | **1** | EG1 | 1.62 | 0.49 | -2.42* |
| | | EG2 | 1.91 | 0.28 | |
| | **2** | EG1 | 1.79 | 0.41 | -1.70* |
| | | EG2 | 1.95 | 0.20 | |
| | **3** | EG1 | 1.66 | 0.48 | -2.11* |
| | | EG2 | 1.91 | 0.28 | |
| **B** | **1** | EG1 | 1.91 | 0.28 | -1.41* |
| | | EG2 | 2 | 0 | |
| | **2** | EG1 | 1.91 | 0.28 | -1.41* |
| | | EG2 | 2 | 0 | |
| | **3** | EG1 | 1.87 | 0.33 | -0.99* |
| | | EG2 | 1.95 | 0.20 | |

We also carried out an emotion analysis in order to measure the children's positive emotions. To that end, we analyzed some photos of the three groups during the experiment. Another instructor (a non-participant observer) was responsible for taking those pictures using a smartphone. He took photos of each student periodically during the 4th week in the middle and at the end of each student's session (about 2.5 and 5 minutes into each turn respectively). A total of 140 photos, 48 of children in the CG, 46 of EG1 and 46 of EG2, were analyzed to identify emotions expressed in the children's faces in order to gather more qualitative information about their levels of satisfaction. Informed consent was required from the children's parents in order to use the photos for the study which is why in Figure 8 the children's eyes are obscured to prevent them from being identified.

In order to analyze emotions through images we used the Microsoft Emotion API (https://www.microsoft.com/cognitive-services/en-us/emotion-api) also known as the Oxford emotion API. We selected this specific API because it can be used to detect not only a face but also a group of faces (the maximum number of faces is 64) and it has been previously used successfully in other educational research (Khalfallah at al. 2017, Saneiro et al. 2014, Takac et al. 2016, Weber et al., 2016, Bharatharaj et al, 2017). The Emotion API uses multilayered deep learning technology to return confidence across a vector of 8 emotions for each face in the image, as well as a bounding box for the face. The emotions detected are anger, contempt, disgust, fear, happiness, neutral, sadness, and surprise. It is necessary to note that this vector provided by the Microsoft Emotion API consisted of eight emotions instead of the seven in Ekman's classic model (Ekman, 1992). This is because of the addition of the contempt emotion. This 8 length emotion vector is also returned by other online facial emotion recognition application programs (Takac et al. 2016). These emotions are understood to be cross-culturally and universally communicated with particular facial expressions. As an example, two images of the same student's turn (EG2 subgroup 1) are shown below (See Figure 8). The image on the left shows children during the session. In the image on the right (at the end of the session) the Emotion API was used to detect faces (indicated by rectangles) and show the emotions of the selected student (the girl in the center of the image).

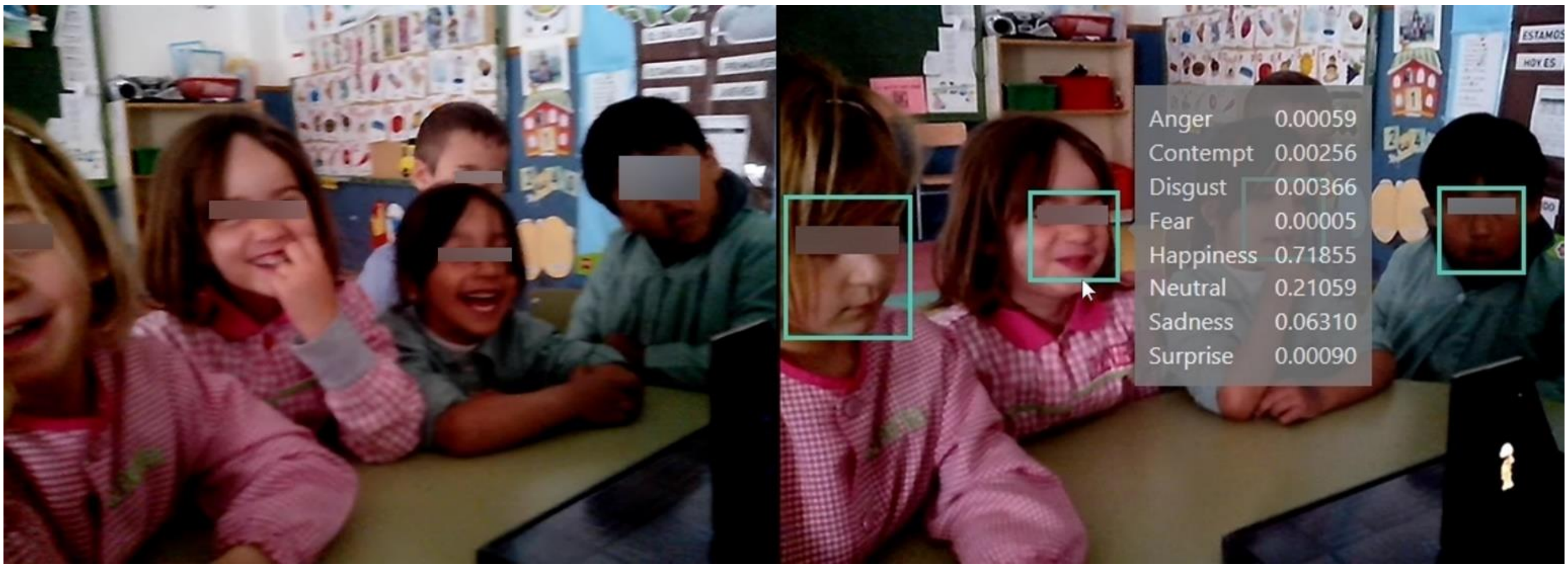

Figure 8: Images of EG2 subgroup 1 children during the experiment.

The Microsoft Emotion API returns a JSON (JavaScript Object Notation) file for each image with the number of faces detected, the coordinates about the location of faces in the image in pixels (left, top, width, and height) and the 8 emotion values (normalized to add up to one). We used this information in order to produce an aggregate result about emotions detected in all photos for each group. Table 7 shows the mean and standard deviation of the emotions detected for each group.

*Table 7.* Descriptive statistics and ANOVA for Children's emotions in CG, EG1 and EG2.

| **Emotion** | **Group** | **Mean** | **SD** | **F statistic** | **Post-hoc Games Howell** |
|---|---|---|---|---|---|
| **Anger** | CG | 0.0030 | 0.0028 | 0.746 | |
| | EG1 | 0.0031 | 0.0034 | | |
| | EG2 | 0.0039 | 0.0046 | | |
| **Contempt** | CG | 0.0007 | 0.0004 | 2.337 | |
| | EG1 | 0.0006 | 0.0003 | | |
| | EG2 | 0.0008 | 0.0004 | | |
| **Disgust** | CG | 0.0008 | 0.0006 | 2.631 | |
| | EG1 | 0.0014 | 0.0021 | | |
| | EG2 | 0.0009 | 0.0012 | | |
| **Fear** | CG | 0.0010 | 0.0022 | 1.624 | |
| | EG1 | 0.0008 | 0.0011 | | |
| | EG2 | 0.0004 | 0.0004 | | |
| **Happiness** | CG | 0.3123 | 0.1788 | 47.359* | EG2>EG1 |
| | EG1 | 0.4804 | 0.1585 | | EG2>CG |
| | EG2 | 0.6190 | 0.1129 | | EG1>CG |
| **Neutral** | CG | 0.6667 | 0.1857 | 58.829* | CG>EG1 |
| | EG1 | 0.4903 | 0.1574 | | CG>EG2 |
| | EG2 | 0.3210 | 0.1089 | | EG1>EG2 |
| **Sadness** | CG | 0.0071 | 0.0053 | 6.636* | CG>EG2 |
| | EG1 | 0.0051 | 0.0046 | | |
| | EG2 | 0.0038 | 0.0029 | | |
| **Surprise** | CG | 0.0085 | 0.0078 | 28.200* | EG2>EG1 |
| | EG1 | 0.0184 | 0.0089 | | EG2>CG |
| | EG2 | 0.0504 | 0.0475 | | EG1>CG |

*p<0.05

Table 7 shows that in general there are two prominent emotions in the three groups, happiness and neutral. However our goal was to check if there were significant differences in the emotions between groups. According to Wilks' Lambda ($\lambda = .375$; $F_{(16,\ 260)}=10.300$; $p<.001$; $\eta^2_{p=}.388$ ) there were differences between groups in that group of emotions. Those differences were found in only four of the emotions: happiness ($F_{(2,\ 137)}=47.359$) $p<.001$; $\eta^2_{p=}.409$), neutral ($F_{(2,\ 137)}=58.829$; $p<.001$; $\eta^2_{p=}.462$), sadness ($F_{(2,\ 137)}=6.636$; $p<.01$; $\eta^2_{p=}.088$), and surprise ($F_{(2,\ 137)}=28.200$; $p<.001$; $\eta^2_{p=}.292$). We performed a multivariate ANOVA and post-hoc analysis to check which groups those differences were between. Prior to that, in order to verify the equality of covariance matrices we carried out Box's Test to test the null hypothesis that the observed covariance matrices of the dependent variables were equal across groups. Results showed that Box's M was statistically significant so we had to assume that there were differences and choose an appropriate post-hoc test. According to Games-Howell there were statistically significant differences between every group in happiness, neutral, and surprise (p<.001), and between CG and EG2 in sadness; mean values were EG2>EG1>CG for happiness and surprise, and CG >EG1>EG2 for neutral and sadness.

Finally, the teacher was briefly interviewed and asked to provide some feedback about the children's behavior during the experiment. He asserted that the children showed a great deal of interest in *Arturito*. He informed us that in general, the children in EG1 and EG2 looked happier in each session and they seemed more focused and engaged than children in the CG. He also stated that the children were excited before the sessions when they knew that they were going to interact with *Arturito*. Above all, most of the EG2 children asked to play with the hologram game again even some weeks after the experience. The teacher also reported that both groups (EG1 and EG2 children) laughed and smiled a lot during all of their sessions with *Arturito*, especially the EG2 children who seemed to be the most excited.

## 5. Discussion and Conclusions

In this study, we adapted a Holographic game in which children have to speak with a virtual teacher called *Arturito*. We developed a specific mobile application using speech recognition to interact with children when practicing pronunciation of English words. The goal of the game was to improve the correct pronunciation of all the words *Arturito* requested in order to see him dancing. It is worth noting that our mobile-based application is not a simple game-based learning application, but an adaptation of a magic trick game. We used a hologram illusion (a realistic 3D image with movement) that appears real; a robot Hologram acting as an instructor using moving 3D images.

To test our hypothesis, we compared three different conditions: a CG corresponding to a traditional teaching methodology; EG1 using a mobile based app that closely mirrored the types of interaction available with the hologram but without the hologram element; and EG2 implementing the holographic technology.

The results from the pre-post-test experiment confirm that the children performed better when practicing pronunciation of basic English vocabulary words using a mobile-based application than when using the traditional methodology. It leads us to believe that that mobile applications can produce better performances than traditional methods, as reported in other related studies with children learning EFL (Liaw, 2014). Furthermore, in addition to the EG2 children having the highest scores of the three groups, there were significant differences compared to the CG but also to the EG1. This may be attributed to the holographic element that could be increasing students' motivation and interest in practicing more than merely mobile-based applications or traditional methods. We have not found any previous work using holograms for teaching or practicing English for children to compare our results to. However, looking back at previous research our results are not surprising, considering the inherent power of games in motivating through fun (Bisson and Luckner 1996) or the results from Mnaathr and Basha (2013) who found that children learned science topics earlier in their educational lives with the application of 3D models.

The results from the satisfaction survey also confirm that the children enjoyed practicing pronouncing English vocabulary using the mobile application developed for this study, and that they preferred that over traditional methods. Children in both EGs were more satisfied with the two versions of *Arturito* than with traditional teaching methods. Additionally, the children in the holographic group scored *Arturito* significantly higher than traditional methods and their scores indicated that they would like him to teach them more "stuff" going beyond the use of images in books and the board. Again, this is not surprising as holographic technology is a novelty that provides a rich variety of graphic representations to generate more realistic scenarios and is literally using technology to represent reality and embody fantasy. The innovation in this case is the application of holographic technology to EFL at an early age.

Moreover, the teacher talked to the children informally about the holographic version of *Arturito* and was given answers like "*Arturito* is alive", "he is real", "it is more fun than the phone", "he is a robot ghost", "I love when he sings and dances", or "it is like magic". This chimes with the idea of students seeking new ways to learn and our perception of the hologram as a new learning tool that increases learning (Golden, 2017). The teacher also informally reported that most of the EG2 children asked to play with the hologram game again even some weeks after the experience, and that some of them, especially those who participate less in traditional learning sessions, interacted more easily with the hologram than with the instructor. Despite this information coming from an informal observation, it does highlight one of the major challenges when intervening in learning, which is the difficulty of maintaining positive intervention effects over time (Melnyk & Morrison-Beedy, 2012).

The teacher also reported that, in general, children seemed very focused, and engaged when using the hologram game. Although this observation would need systematization and further investigation, this preliminary evidence is consistent with previous research (Annetta et al., 2009). The most remarkable gains that the teacher reported were in terms of attention, engagement, and curiosity. The teacher also stated that the hologram kept those children that have some attention difficulties in the regular learning sessions focused on the task which is in line with results from Durango (2015). It is not surprising to achieve attentional outcomes when using games for learning (Boyle et al., 2016). Children learn through playing games because this is the easiest way for them to connect with objects and the world around them, so whenever a game is introduced in language classes, students' motivation to pay attention is thought to increase (Yiltanhhlar & Kivanc, 2015).

In addition, the art of magic has the potential to amaze and capture and hold the attention of people of all ages (Spencer, 2012). The analysis of emotions in photographs taken of the children during the sessions suggested that children using the application with the holographic game exhibited more happiness and surprise, and less sadness and neutral emotion, than those using only the smartphone or the traditional teaching methods. It is easy to guess the effect of happiness on learning (Csikszentmihalyi, 2014) in contrast to emotions identified in previous research as low in arousal such as neutral feelings and boredom (Harley, Bouchet, Hussain, Azevedo, & Calvo, 2015). It is much more challenging to reflect on previous results showing that positive emotions foster academic achievement only when they are mediated by motivation (Mega, Ronconi, De Beni, 2014). Previous research has also shown that holograms seem to arouse curiosity in children, and people are better at learning information that they are curious about (Gruber, Gelman, & Ranganath, 2014). Although it may sound banal, this emotional-motivational state plays an important role when acquiring knowledge and making the process pleasurable (D'Mello, 2012; Litman, 2015). Other results have also shown that academic emotions are significantly related to students' motivation, learning strategies, cognitive resources, self-regulation, and academic achievement (Pekrun, Goetz, Titz, & Perry, 2002). The findings indicate that affective research in educational and computer science should acknowledge the role of emotional importance in academic settings by addressing the study of the full range of emotions experienced by students when introducing cutting edge educational methods. It is interesting to note that we only took photographs in the final session (week four) in order to avoid the effect of novelty from the technology. The users of mobile, technology games, and other types of computer-based instruction experience a transient effect because of their novelty (Kulik and Kulik, 1991).

Much more research needs to be conducted before concluding that a hologram is better than a teacher or can substitute for a teacher, something that is not on the horizon of our current work, but this approach could be part of a radical change in instructional style based on ICTs applied to education. The greatest promise of educational software is not teaching efficiency but motivating students and improving their learning experience, and the core contribution is not so much the teaching as the child's learning process. Introducing tools such as *Arturito* in educational settings could reduce practicing time and instructor load, affording opportunities for drill and practice, for example. Additionally, these kinds of technologies enable students to be more responsible and self-regulated, and this could be a help to teachers who have been trying to achieve this since the constructivist approach to learning emerged (Fat, 2000). In fact, a holographic resource could be understood as a cutting-edge evolution of the already traditional means such as blackboards, text books, playing cards, and interactive whiteboards.

The potential of a projected 3D image in technical (portability, personalization, etc.) and learning terms (novelty, gamification, etc.) is extremely promising. In fact, life-size holographic images that speak and answer questions can be used to enhance the educational process by bringing famous characters from the past back to life, to speak about themselves, to explain something as an assistant teacher (Ghuloum, 2010), as a teaching aid to prompt viewers (Walker, 2012), or basically as a general training aid for different specializations (Leuski et al., 2006). Holograms have already been used successfully in this sense in adult learning environments, especially medicine. Medical Holography for Basic Anatomy Training allowed learners to observe full parallax, auto-stereoscopic 3D human anatomy images first hand (Hackett, 2013). Holograms of human organs have also been used to teach medical students topics ranging from simple dissections and health protocols to the latest surgical techniques (Ko, 1998). Holograms have also uniquely facilitated the spontaneous understanding of human neuroanatomical relationships that cannot be easily learned using photographs or diagrams (Ko and Webster, 1995). In other domains, holography provides better visual aids than either photographs or line drawings in training in domains such as construction, technical documentation, and storage (Frey and Eichert, 1978), and provides better perception of the 3D model shapes for mechanical engineering parts when learning how to draw them (Figueiredo et al., 2014).

Our study is subject to some limitations. We are conscious of the assessment methodology used, the performance tests only inform us about the children's declarative knowledge. In addition, a self-informed survey to measure satisfaction and the methodology preference in the sample is only valid as a preliminary result. Considering the potential of holograms for practice, it would be interesting to systematically explore other core variables in learning such as metacognition, self-regulation, or the motivational spectrum. Aspects for providing a more personalized gaming experience to sustain the engagement of the players with the game may also be considered. For that reason, one of our future plans is to implement observational video tools in order to test and improve the learning experience. While there are methodological difficulties with this, an observational video instrument would allow us to make more valid assessments of the children's cognitive and metacognitive processes during learning with the hologram, and not be restricted to just the result. Currently, the work from Whitbread and colleagues is heading in this direction, constructing and developing observational instruments from recording early aged child 'events' that are very helpful to our approach (Whitebread et al., 2009; Whitebread and Coltman, 2015; Whitebread et al., 2009).

Furthermore, the Flow Theory approach will undoubtedly improve the experience when practicing, particularly in the sequential and random modes. Flow theory is a new subject worth considering in EFL learning (Guan, 2013). An interesting future prospect for this research is to dynamically adapt the game to balance it to each child's skill level and the challenge of the task. When a task is too difficult, it might make the child anxious, when a task is too easy, that may end up causing inattentiveness or boredom. When the task is just right, children are focused and immersed in the learning process, a state of flow that seems to be useful for sustaining players' engagement with the game (Sajjadi, Van Broeckhoven, & De Troyer, 2014).

Another limitation is the quasi-experimental nature of using existing classes. The next steps are to replicate the experiment using a strict experimental design, in which children would be randomly assigned to classes. We also want to increase the English skills contained in the pre-school curricula. In the future, we want also to use *Arturito* not only for practicing and revising the pronunciation of some previously learned words, but also for other tasks such as teaching new vocabulary to children. We aim to have *Arturito* introduce the new words from a specific domain or context in a similar way to instructors. However, one must remember that computer games for learning are particularly effective when addressing a certain skill (Griffiths 2002), for example in encouraging learning in curriculum areas such as maths, physics and languages, where specific objectives can be stated (Randel, Morris, Wetzel, & Whitehill 1992). We also want to increase the difficulty of the learning material in an attempt to gather more information about instructional practice. As we saw in the results, the children's pre-test scores were already quite good for every group, so the potential progress that could be achieved by the EGs is smaller and harder to achieve than if the margin were bigger. The good news is that with a simple software improvement, the student can be presented with almost limitless content and be given differing levels of challenge, tasks can be instantly updated, customized and modified by teachers or even individual players, so that the player becomes part of the creative team.

Another limitation which is unfortunately typical in this kind of work is that we do not know how long our intervention results are sustained for. However, this is not a core aspect in our study as we are not proposing a remedial intervention but a tool to add to the practicing-learning routine. Moreover, we conducted the post-test one week after the experiments trying to test long-term memory and avoiding immediate transience, or forgetting that occurs with the passage of time. In the future, it would be necessary to take repeated measures to test the effect of the hologram on both practicing and levels of motivation over time. In a large number of similar studies on EFL vocabulary learning, post-tests were administered immediately after experiments (Agca and Ozdemir, 2013, Ashraf et al. 2015, Mashhadi and Jamalifar, 2015). Raiche (2011) proposed waiting at least 24 hours or ideally a few days to test long-term retention but only a proportion of research into second language vocabulary learning administered multiple post-tests. Barcroft (2007) conducted two post-tests, two days later and one week later to check if the effect was maintained over time, and in a related study by Schuetze (2017), three retention tests were carried out: one day after the last practice, four weeks after the last practice, and eight weeks after the last practice. This kind of tracking is also supported by language instructors (Oxford, 1990).

Finally, with respect to the mobile application and hologram, some technical improvements are already being made to improve the learning experience; the size of the hologram and the volume of the hologram's voice. Currently, the hologram could be bigger and its voice louder, allowing larger student groups. We could use a tablet instead of a smartphone and we could use an additional speaker. These simple improvements would make the experience smoother in a class/group with more students. In conclusion, we recognize that mobile-based applications and holograms are not a panacea but if innovative technologies are more engaging and appealing to students and if, in turn, these learners are more motivated to interact with these learning environments than with traditional materials, then this in itself may justify the use of and deeper research into these resources.

## Acknowledges

The authors gratefully acknowledge the financial subsidy provided by Spanish Ministry of Science and Technology TIN2017-83445-P, EDU2014-57571-P and the Principality of Asturias, through its Science, Technology and Innovation Plan (GRUPIN14-053).